\documentclass[prd,onecolumn,nofootinbib]{revtex4-2}

\usepackage[utf8]{inputenc}
\usepackage{lmodern}
\pdfoutput=1
\usepackage{amsmath,amssymb}
\usepackage{graphicx}
\usepackage{hyperref}
\usepackage{cmap}

\begin{document}

\title{The Scalar Mach-Sciama Theory of Gravitation}

\author{Vel\'asquez-Toribio, A. M.}
\affiliation{Center for Astrophysics and Cosmology, Federal University of Esp\'irito Santo, 29075-910 Vit\'oria - ES, Brazil}
\date{\today}

\begin{abstract}
We formulate a scalar realization of Sciama’s Machian programme within the general
Bergmann--Wagoner class of scalar--tensor gravity. Starting from a universally conformally coupled
matter sector, we rewrite the field equations in terms of the invariant set
$\{{\cal I}_1,{\cal I}_2,{\cal I}_3,\hat g_{\mu\nu}\}$, so that Machian requirements can be stated
independently of conformal frame. Sciama’s causal postulate is implemented not by modifying the
local dynamics, but as a selection rule on the solution space of the invariant scalar equation:
the admissible configuration is the retarded response to the matter distribution in the causal
past, with any source-free contribution removed. In a spatially flat FLRW universe, and in the
light, slowly varying regime, the prescription reduces to an explicit temporal kernel that links
the background scalar evolution to the matter content within the Hubble region, reproducing the
expected Machian scaling in an expanding background. Universal coupling to a single physical metric implies that structureless test bodies share the
same acceleration in a given external configuration, so that the E\"otv\"os parameter vanishes at
leading order. Nonuniversal effects can arise only when gravitational binding energy contributes
appreciably to the total mass, as in strongly self-gravitating objects. Thus, the theory implements
a causal Machian determination of the cosmological inertial scale while remaining consistent with
standard weak-field tests.
\end{abstract}

\maketitle

\section{Introduction}

The notion that inertia should be determined by the rest of the universe has a long intellectual history that predates modern field theory. In early modern philosophy, Leibniz argued against Newtonian absolutism by defending a relational conception of space and motion, articulated most clearly in the Leibniz--Clarke correspondence (1715--1716) \cite{LeibnizClarke1715}. Berkeley further sharpened the operational critique of absolute space in \emph{De Motu} (1721) \cite{Berkeley1721}, and his dictum \textit{esse est percipi}, "to be is to be perceived", captures the demand that fundamental notions be grounded in observables \cite{Berkeley1710}.

Mach recast these philosophical objections as a physical programme: inertial properties are not intrinsic attributes of bodies, but reflect their dynamical relations to the total mass--energy content of the universe \cite{Mach1883}. Einstein initially hoped that General Relativity (GR) would provide a complete realization of this viewpoint. He later stressed, however, that ``Mach's principle'' should not be conflated with general covariance, and he also noted that GR admits nontrivial vacuum solutions in which spacetime geometry (and thus inertial structure) exists without matter \cite{Einstein1916GR,Einstein1916Mach}. This limitation has motivated many subsequent attempts to formulate a sharper, explicitly relativistic version of Machian ideas 
\cite{Einstein1918,SEPGenRelEarly,BondiSamuel1964,Raine1981,BarbourBertotti1982,BarbourPfister1995,Ellis2001,
LichteneggerMashhoon2004, Gine2006,Barbour2010Definition,Licata2014, Kragh2012,Fay2024Sciama,Fay2024MachHypotheses}.

In this setting, Brans and Dicke proposed a scalar-tensor theory in which gravity is mediated not only by the metric but also by a scalar field whose dynamics can be sensitive to the cosmic mass--energy distribution \cite{BransDicke1961}. In the Brans--Dicke framework the gravitational ``constant'' becomes dynamical, and the strength of gravity can, in principle, track large-scale properties of the universe. For this reason, modern discussions of Mach's principle in relativistic field theory are often framed within the Brans--Dicke family and its general scalar--tensor extensions.

A distinct and particularly suggestive proposal was put forward by Sciama, who modeled inertia by analogy with Maxwellian electrodynamics. In Sciama's original formulation, inertial reaction forces arise from retarded interactions with distant matter, in direct analogy with the acceleration-dependent part of the electromagnetic Liénard--Wiechert fields \cite{Sciama1953,Sciama1959,Sciama1964}. He later emphasized that a consistent relativistic implementation of Machian boundary conditions is naturally expressed in terms of retarded (causal) Green functions and integral formulations of the field equations \cite{Sciama1964}.

In this paper we keep Sciama's guiding motivation but implement it through a scalar mechanism: a scalar-tensor theory evolving on a spatially flat FLRW background. The scalar sector provides a clean and manifestly covariant setting in which causal propagation is controlled by a retarded Green function, and in which the influence of the cosmic matter distribution on local inertial properties can be made explicit. In particular, the inertial calibration is encoded through a field-dependent mass scale of the form $m(\phi)=m_0 A(\phi)$.

This paper is organized as follows. In Sec.~II we introduce the scalar--tensor action with a universal conformal coupling to matter, derive the field equations, and define the frame-invariant quantities used throughout. In Sec.~III we construct the retarded Green-function solution on an FLRW background in terms of these invariants, and show how it fixes the emergence of inertial mass. In Sec.~IV we present the induced-gravity mechanism that allows us to exclude de~Sitter-type solutions as inertial vacua. In Sec.~V we discuss the equivalence principle in the test-body limit, and in Sec.~VI we summarize our conclusions and outline future directions.

\section{The basic and frame-invariant equations}
\label{sec:basic_eqs_frame_invariants}

We consider the Bergmann--Wagoner class of scalar--tensor theories, i.e.\ models
without higher-derivative terms or derivative--curvature couplings beyond
$(\nabla\Phi)^2$ \cite{Bergmann1968,Wagoner1970,FaraoniBook2004}. We adopt the
metric signature $(-,+,+,+)$ and units $c=1$. The dynamical fields are a
spacetime metric $g_{\mu\nu}$ and a real scalar field $\Phi$. Matter fields $\psi$
couple minimally to a physical (matter) metric $\tilde g_{\mu\nu}$, which we
assume to be conformally related to $g_{\mu\nu}$ by a universal coupling function
$\alpha(\Phi)$. The action is
\begin{eqnarray}
S[g_{\mu\nu},\Phi,\psi]
&=&
\int d^4x\,\sqrt{-g}\,
\left[
\frac{1}{2}\,A(\Phi)\,R
-
\frac{1}{2}\,B(\Phi)\,g^{\mu\nu}\nabla_\mu\Phi\nabla_\nu\Phi
-
U(\Phi)
\right]
\;+\;
S_m\!\left[\tilde g_{\mu\nu},\psi\right],
\label{eq:ST_action_general}
\\
\tilde g_{\mu\nu}
&=&
e^{2\alpha(\Phi)}\,g_{\mu\nu}.
\label{eq:matter_metric_def}
\end{eqnarray}
This parametrization is standard in the scalar-tensor literature (up to
conventional normalizations and minor notational changes), see
e.g.\ Refs.~\cite{FujiiMaeda2003, FaraoniBook2004, DamourEspositoFarese1996BinaryPulsar}.
It contains four a priori arbitrary functions of $\Phi$. The curvature coupling
$A(\Phi)$ multiplies the Ricci scalar and controls the effective gravitational
scale (equivalently, the normalization of the spin-2 kinetic term). The function
$B(\Phi)$ provides a noncanonical kinetic coupling for $\Phi$, while $U(\Phi)$ is
the scalar self-interaction potential. Furthermore, $\alpha(\Phi)$ characterizes
the universal conformal coupling between the gravitational metric $g_{\mu\nu}$ and
the physical metric $\tilde{g}_{\mu\nu}$ governing matter dynamics. A specific
theory within this class is uniquely defined by the choice of $A, B, U$, and
$\alpha$. 

It is sometimes convenient to display an overall gravitational prefactor, often
written as $1/2\kappa$ or $1/2\kappa^2$. In the form (\ref{eq:ST_action_general})
this constant has been absorbed into the definition of $A(\Phi)$ (and likewise
into $B(\Phi)$ and $U(\Phi)$). Concretely, one may restore it by the purely
conventional rescaling
\begin{eqnarray}
A(\Phi)=\frac{F(\Phi)}{\kappa^2},
\qquad
B(\Phi)=\frac{Z(\Phi)}{\kappa^2},
\qquad
U(\Phi)=\frac{V(\Phi)}{\kappa^2}.
\label{eq:kappa_absorption}
\end{eqnarray}

In what follows we use "Jordan frame" in the operational (matter-frame) sense
that matter is minimally coupled to $\tilde g_{\mu\nu}$ and therefore obeys the
standard covariant conservation law with the Levi--Civita derivative of
$\tilde g_{\mu\nu}$. The metric $g_{\mu\nu}$ is not assumed to be physical; it is
an arbitrary representative used to parametrize the same underlying theory. This
also explains why two conformal functions appear: $\alpha(\Phi)$ fixes the
physical matter coupling in (\ref{eq:matter_metric_def}), while the
representative change (\ref{eq:frame_transforma}) below is a field redefinition of
the gravitational variables. Below, Sciama's causal selection rule will be
written as a condition on $\delta{\cal I}_3$ and $\hat g_{\mu\nu}$, avoiding
frame-dependent statements about $A(\Phi)$ or $\Phi$ themselves.

Varying the action with respect to $g_{\mu\nu}$ yields the metric field equation
\begin{eqnarray}
A(\Phi)\,G_{\mu\nu}
&=&
T_{\mu\nu}
+
\nabla_\mu\nabla_\nu A(\Phi)
-
g_{\mu\nu}\,\Box A(\Phi)
\nonumber\\
&&
+
B(\Phi)\left(
\nabla_\mu\Phi\nabla_\nu\Phi
-
\frac{1}{2}\,g_{\mu\nu}(\nabla\Phi)^2
\right)
-
g_{\mu\nu}\,U(\Phi),
\label{eq:metric_field_eq}
\end{eqnarray}
where $\Box\equiv g^{\mu\nu}\nabla_\mu\nabla_\nu$, $(\nabla\Phi)^2\equiv
g^{\mu\nu}\nabla_\mu\Phi\nabla_\nu\Phi$, and $\nabla$ is the Levi--Civita
derivative compatible with $g_{\mu\nu}$. The matter stress tensor associated with
the representative metric $g_{\mu\nu}$ is defined by
\begin{eqnarray}
T_{\mu\nu}
\equiv
-\frac{2}{\sqrt{-g}}\,
\frac{\delta S_m}{\delta g^{\mu\nu}},
\label{eq:stress_energy_def}
\end{eqnarray}
with trace $T\equiv g^{\mu\nu}T_{\mu\nu}$.

Varying with respect to $\Phi$ gives the scalar field equation
\begin{eqnarray}
B(\Phi)\,\Box\Phi
+
\frac{1}{2}\,B_{,\Phi}(\Phi)\,(\nabla\Phi)^2
+
\frac{1}{2}\,A_{,\Phi}(\Phi)\,R
-
U_{,\Phi}(\Phi)
&=&
-\alpha_{,\Phi}(\Phi)\,T.
\label{eq:scalar_field_eq}
\end{eqnarray}
(We omit standard intermediate variation identities; see, e.g., Wald \cite{Wald1984}.)

Taking the trace of Eq.~(\ref{eq:metric_field_eq}) yields
\begin{eqnarray}
-A(\Phi)\,R
=
T
-
3\,\Box A(\Phi)
-
B(\Phi)\,(\nabla\Phi)^2
-
4\,U(\Phi),
\label{eq:trace_metric_eq}
\end{eqnarray}
which is useful when eliminating $R$ between (\ref{eq:metric_field_eq}) and
(\ref{eq:scalar_field_eq}).

Because matter is minimally coupled to the physical metric $\tilde g_{\mu\nu}$,
its equations of motion imply covariant conservation with respect to the
$\tilde g_{\mu\nu}$-compatible derivative $\tilde\nabla$,
\begin{eqnarray}
\tilde\nabla_\mu\,\tilde T^{\mu}{}_{\nu}=0,
\label{eq:tilde_conservation}
\end{eqnarray}
where the physical (matter-frame) stress tensor is
\begin{eqnarray}
\tilde T_{\mu\nu}
\equiv
-\frac{2}{\sqrt{-\tilde g}}\,
\frac{\delta S_m}{\delta \tilde g^{\mu\nu}}.
\label{eq:Ttilde_def}
\end{eqnarray}
For $\tilde g_{\mu\nu}=e^{2\alpha}g_{\mu\nu}$ one obtains the standard conformal
relations between the two stress tensors,
\begin{eqnarray}
T_{\mu\nu}
=
e^{2\alpha}\,\tilde T_{\mu\nu},
\qquad
T=e^{4\alpha}\,\tilde T,
\label{eq:T_Ttilde_relation}
\end{eqnarray}
so the matter source term in Eq.~(\ref{eq:scalar_field_eq}) can be written in
either form. In contrast to (\ref{eq:tilde_conservation}), the tensor $T_{\mu\nu}$
defined with respect to $g_{\mu\nu}$ is generally not conserved under $\nabla$
when $\alpha_{,\Phi}\neq 0$; using the field equations one finds the exchange law
\begin{eqnarray}
\nabla^\mu T_{\mu\nu}
=
-\,\alpha_{,\Phi}(\Phi)\,T\,\nabla_\nu\Phi.
\label{eq:nonconservation_Einsteinframe}
\end{eqnarray}

A further freedom, independent of the physical matter coupling encoded in
$\alpha(\Phi)$, is to change the representative variables $(g_{\mu\nu},\Phi)$ by
a conformal rescaling of the representative metric combined with a scalar-field
reparametrization,
\begin{eqnarray}
g_{\mu\nu}
=
e^{2\gamma(\bar\Phi)}\,\bar g_{\mu\nu},
\qquad
\Phi
=
f(\bar\Phi),
\label{eq:frame_transforma}
\end{eqnarray}
with $e^{2\gamma(\bar\Phi)}>0$ and $f$ monotonic. The standard conformal
transformation rules,
\begin{eqnarray}
R[g]
=
e^{-2\gamma}\Bigl(
\bar R
-
6\,\bar\Box\gamma
-
6\,(\bar\nabla\gamma)^2
\Bigr),
\qquad
\sqrt{-g}=e^{4\gamma}\sqrt{-\bar g},
\label{eq:conformal_R_and_det}
\end{eqnarray}
lead to the transformation of the defining functions after integrating by parts
the $\bar\Box\gamma$ term. In particular, the physical metric can be written
equivalently as
\begin{eqnarray}
\tilde g_{\mu\nu}
=
e^{2\alpha(\Phi)}g_{\mu\nu}
=
e^{2\bar\alpha(\bar\Phi)}\,\bar g_{\mu\nu},
\qquad
\bar\alpha(\bar\Phi)=\alpha(f(\bar\Phi))+\gamma(\bar\Phi),
\label{eq:physical_metric_barred}
\end{eqnarray}
and the action retains the form of Eq.~(\ref{eq:ST_action_general}) but with new
functions $\bar A(\bar\Phi)$, $\bar B(\bar\Phi)$, $\bar U(\bar\Phi)$:
\begin{eqnarray}
\bar A(\bar\Phi)
=
e^{2\gamma(\bar\Phi)}\,A\!\left(f(\bar\Phi)\right),
\qquad
\bar U(\bar\Phi)
=
e^{4\gamma(\bar\Phi)}\,U\!\left(f(\bar\Phi)\right),
\qquad
\bar\alpha(\bar\Phi)=\alpha(f(\bar\Phi))+\gamma(\bar\Phi),
\label{eq:transform_A_U_alpha}
\\
\bar B(\bar\Phi)
=
e^{2\gamma(\bar\Phi)}
\left[
\left(f'(\bar\Phi)\right)^2\,B\!\left(f(\bar\Phi)\right)
-
6\left(\gamma'(\bar\Phi)\right)^2\,A\!\left(f(\bar\Phi)\right)
-
6\,\gamma'(\bar\Phi)\,f'(\bar\Phi)\,A_{,\Phi}\!\left(f(\bar\Phi)\right)
\right],
\label{eq:transform_B}
\end{eqnarray}
where $f'\equiv d\Phi/d\bar\Phi$ and $\gamma'\equiv d\gamma/d\bar\Phi$.

An Einstein-frame representative is obtained by choosing $\gamma$ so that $\bar A$
is constant; one convenient choice is
\begin{eqnarray}
e^{2\gamma(\bar\Phi)}=\frac{1}{A(f(\bar\Phi))},
\qquad\Rightarrow\qquad
\bar A(\bar\Phi)=1,
\qquad
\bar g_{\mu\nu}=A(\Phi)\,g_{\mu\nu}.
\label{eq:Einstein_choice_gamma}
\end{eqnarray}

A convenient representation-independent set of invariants was proposed by
J\"arv, Kuusk, Saal and Vilson
\cite{JarvKuuskSaalVilson2015Invariant,JarvKuuskSaalVilson2015Transform}. The
first two are immediate consequences of (\ref{eq:physical_metric_barred}) and
(\ref{eq:transform_A_U_alpha}),
\begin{eqnarray}
\frac{e^{2\bar\alpha(\bar\Phi)}}{\bar A(\bar\Phi)}
=
\frac{e^{2\alpha(\Phi)}}{A(\Phi)},
\qquad
\frac{\bar U(\bar\Phi)}{\bar A(\bar\Phi)^2}
=
\frac{U(\Phi)}{A(\Phi)^2}.
\label{eq:I1I2_quick_check}
\end{eqnarray}
The third invariant follows from the exact identity (see
Refs.~\cite{JarvKuuskSaalVilson2015Invariant, JarvKuuskSaalVilson2015Transform,KaramTamvakis2018FrameInvariant,Dicke1962,KaramTamvakis2019Inflation})
\begin{eqnarray}
2\bar A\bar B + 3\left(\bar A_{,\bar\Phi}\right)^2
=
e^{4\gamma}\,\left(f'\right)^2\,\left(2AB+3A_{,\Phi}^2\right),
\label{eq:key_identity_I3}
\end{eqnarray}
which implies the invariance of the field-space line element
\begin{eqnarray}
\left(d{\cal I}_3\right)^2
=
\frac{2AB+3A_{,\Phi}^2}{4A^2}\,d\Phi^2
=
\frac{2\bar A\bar B+3\bar A_{,\bar\Phi}^2}{4\bar A^2}\,d\bar\Phi^2.
\label{eq:I3_line_element_invariance}
\end{eqnarray}

With these motivations, the standard set of three basic invariants is
\begin{eqnarray}
{\cal I}_1(\Phi)
\equiv
\frac{e^{2\alpha(\Phi)}}{A(\Phi)},
\label{eq:I1_def}
\qquad
{\cal I}_2(\Phi)
\equiv
\frac{U(\Phi)}{A(\Phi)^2},
\label{eq:I2_def}
\\
\left(\frac{d{\cal I}_3}{d\Phi}\right)^2
\equiv
\frac{2\,A(\Phi)\,B(\Phi)+3\,[A_{,\Phi}(\Phi)]^2}{4\,A(\Phi)^2},
\label{eq:I3_def}
\end{eqnarray}
together with the invariant metric
\begin{eqnarray}
\hat g_{\mu\nu}
\equiv
A(\Phi)\,g_{\mu\nu}.
\label{eq:Einstein_invariant_metric}
\end{eqnarray}
Finally, combining (\ref{eq:matter_metric_def}) and (\ref{eq:Einstein_invariant_metric})
gives the physical metric directly in terms of invariant variables,
\begin{eqnarray}
\tilde g_{\mu\nu}
=
e^{2\alpha}g_{\mu\nu}
=
\frac{e^{2\alpha}}{A}\,\hat g_{\mu\nu}
=
{\cal I}_1(\Phi)\,\hat g_{\mu\nu}.
\label{eq:tildeg_in_terms_of_I1_ghat}
\end{eqnarray}
Statements formulated purely in terms of ${\cal I}_1$, ${\cal I}_2$, ${\cal I}_3$
and $\hat g_{\mu\nu}$ are therefore independent of the representative choice
$(g_{\mu\nu},\Phi)$. In particular, ${\cal I}_1$ tracks the relative
normalization between matter units and the gravitational sector, ${\cal I}_2$
packages the self-interaction potential in invariant gravitational units, and
${\cal I}_3$ provides an invariant canonical scalar variable. These objects will
be used below to express Mach-type restrictions and the retarded selection rule
as conditions on $\delta{\cal I}_3$ and $\hat g_{\mu\nu}$, rather than on
frame-dependent quantities such as $A(\Phi)$ or $\Phi$.

\section{The Sciama--Mach Model}
\label{sec:sciama_mach}

Section~\ref{sec:basic_eqs_frame_invariants} provides a fundamental demarcation between 
two independent theoretical structures: (i) the universal matter coupling, which 
characterizes the physical metric $\tilde{g}_{\mu\nu}$ through Eq.~(\ref{eq:matter_metric_def}), 
and (ii) the covariance under field redefinitions $(g_{\mu\nu}, \Phi) \mapsto (\bar{g}_{\mu\nu}, \bar{\Phi})$ 
that leave the scalar--tensor model unchanged. Building upon this structural 
independence, we introduce Sciama's causal (Machian) selection as an additional 
physical postulate.

Ensuring that this selection rule remains decoupled from any specific conformal 
representation requires its direct implementation within the frame-invariant 
framework, defined by the invariants $\mathcal{I}_1, \mathcal{I}_2, \mathcal{I}_3$ 
and the invariant metric $\hat{g}_{\mu\nu}$. Such a representation-independent 
approach, following the methodology established in Ref.~\cite{JarvKuuskSaalVilson2015Invariant}, 
reformulates the scalar-field equation strictly in terms of these invariant 
variables. Consequently, a convenient invariant form of the scalar equation 
is expressed as \cite{,JarvKuuskSaalVilson2015Invariant, JarvKuuskSaalVilson2015Transform,KaramTamvakis2018FrameInvariant,Dicke1962,KaramTamvakis2019Inflation}
\begin{eqnarray}
\hat\square\,\mathcal{I}_3
-
\frac{1}{2}\,\frac{d\mathcal{I}_2}{d\mathcal{I}_3}
=
-\frac{1}{4}\,\frac{d\ln\mathcal{I}_1}{d\mathcal{I}_3}\,\hat T,
\label{eq:I3_invariant_eom_SIII}
\end{eqnarray}
where $\hat\square\equiv \hat g^{\mu\nu}\hat\nabla_\mu\hat\nabla_\nu$ is the d'Alembertian constructed
from the invariant metric $\hat g_{\mu\nu}$ and its Levi--Civita derivative $\hat\nabla_\mu$.
Moreover, $\mathcal{I}_2(\Phi)\equiv U(\Phi)/A(\Phi)^2$ is the invariant potential, and $\hat T$ denotes
the trace of the matter stress tensor expressed in invariant gravitational units. Defining
\begin{eqnarray}
\hat T_{\mu\nu}\equiv \frac{1}{A(\Phi)}\,T_{\mu\nu},
\qquad
\hat T\equiv \hat g^{\mu\nu}\hat T_{\mu\nu},
\label{eq:That_def_SIII}
\end{eqnarray}
one obtains the exact relation
\begin{eqnarray}
\hat T=\frac{T}{A(\Phi)^2},
\label{eq:That_vs_T_SIII}
\end{eqnarray}
which will be used repeatedly below.

Sciama's causal requirement is not implemented by modifying the local field equation
(\ref{eq:I3_invariant_eom_SIII}) or by deforming its differential operator. The local operator is
completely fixed once the theory and the invariant metric $\hat g_{\mu\nu}$ are specified. What is
imposed instead is a physical selection criterion on the space of solutions: among all solutions of
the same local hyperbolic equation, we retain only those for which the scalar configuration relevant
for inertial calibration is generated by the matter distribution in the causal past. In particular,
no independent source-free (homogeneous) mode is allowed to contribute to the physically admissible
configuration.
To implement Sciama's causal postulate in a controlled setting, we work at the level of linear
response around a cosmological background. We split the invariant scalar into a homogeneous FLRW
background and a perturbation,
\begin{eqnarray}
\mathcal{I}_3(x)=\bar{\mathcal{I}}_3(t)+\delta\mathcal{I}_3(x),
\label{eq:I3_split_background_pert_SIII}
\end{eqnarray}
where $\bar{\mathcal{I}}_3(t)$ solves the background (unperturbed) cosmological problem, while
$\delta\mathcal{I}_3(x)$ captures the spacetime-dependent deviation induced by inhomogeneous matter
sources and local departures from the background configuration.

At fixed $\bar{\mathcal{I}}_3$, the perturbation obeys a linear equation of the schematic form
\begin{eqnarray}
\hat{\mathcal{D}}_x\,\delta\mathcal{I}_3(x)=\hat{\mathcal{S}}(x),
\label{eq:operator_form_SIII}
\end{eqnarray}
where $\hat{\mathcal{D}}$ is a normally hyperbolic operator constructed from $\hat\square$ together
with an effective mass term determined by the background, for instance through
$\left.(d^2\mathcal{I}_2/d\mathcal{I}_3^2)\right|_{\bar{\mathcal{I}}_3}$. The corresponding
linearized source can be written as
\begin{eqnarray}
\hat{\mathcal{S}}(x)
=
-\frac{1}{4}
\left.\frac{d\ln\mathcal{I}_1}{d\mathcal{I}_3}\right|_{\bar{\mathcal{I}}_3}\,
\delta\hat T(x),
\label{eq:source_def_SIII}
\end{eqnarray}
with the understanding that one may equivalently replace $\delta\hat T$ by $\hat T$ when the
background trace vanishes.

Let $\hat G_{\rm ret}(x,x')$ denote the retarded Green function associated with $\hat{\mathcal{D}}$,
defined by
\begin{eqnarray}
\hat{\mathcal{D}}_x\,\hat G_{\rm ret}(x,x')
=
\frac{\delta^{(4)}(x-x')}{\sqrt{-\hat g(x)}},
\qquad
\hat G_{\rm ret}(x,x')=0
\ \ \text{if}\ \ x'\notin J^{-}(x),
\label{eq:Green_def_SIII}
\end{eqnarray}
where $J^{-}(x)$ is the causal past of $x$ with respect to $\hat g_{\mu\nu}$, i.e.\ the set of points
$x'$ that can reach $x$ along future-directed timelike or null curves. Since $\tilde g_{\mu\nu}$ and
$\hat g_{\mu\nu}$ are conformally related, they share the same null cones; consequently, the notion
of retarded support in Eq.~(\ref{eq:Green_def_SIII}) is invariantly defined.

Because Eq.~(\ref{eq:operator_form_SIII}) is linear, its general solution decomposes uniquely into a
sourced contribution plus a source-free contribution,
\begin{eqnarray}
\delta\mathcal{I}_3(x)=\delta\mathcal{I}_3^{\rm ret}(x)+\delta\mathcal{I}_3^{\rm free}(x),
\qquad
\hat{\mathcal{D}}\,\delta\mathcal{I}_3^{\rm free}(x)=0,
\label{eq:deltaI3_ret_plus_free}
\end{eqnarray}
where ``free'' refers to a solution of the homogeneous (source-free) linearized equation, not to
spatial homogeneity in the FLRW sense. A convenient representative of the sourced sector is the
retarded particular solution,
\begin{eqnarray}
\delta\mathcal{I}_3^{\rm ret}(x)
=
\int d^4x'\,\sqrt{-\hat g(x')}\;
\hat G_{\rm ret}(x,x')\;
\hat{\mathcal{S}}(x').
\label{eq:Sciama_retarded_rule_SIII}
\end{eqnarray}
By construction, $\delta\mathcal{I}_3^{\rm ret}(x)$ depends only on sources within $J^{-}(x)$.

Sciama's causal (Machian) selection rule is the additional physical requirement that the
configuration relevant for inertial calibration contains no independent free component. In the
present linearized setting this is implemented by imposing
\begin{eqnarray}
\delta\mathcal{I}_3^{\rm free}(x)=0,
\qquad\Rightarrow\qquad
\delta\mathcal{I}_3(x)=\delta\mathcal{I}_3^{\rm ret}(x).
\label{eq:Sciama_free_zero}
\end{eqnarray}
The motivation is Machian: $\delta\mathcal{I}_3^{\rm free}$ encodes independent scalar initial data
that are not fixed by the matter distribution and would therefore introduce an intrinsic inertial
calibration unrelated to cosmic sources. The retarded selection rule removes this arbitrariness and
enforces instead that the scalar perturbation relevant for inertia is entirely generated by past
matter sources and propagated causally according to the same local field equation \footnote{%
This remark makes explicit that the Machian postulate can be stated either as a choice of the
retarded Green function or, equivalently, as a condition on initial data. For a normally hyperbolic
operator on a globally hyperbolic spacetime, the Cauchy problem is well posed: given data on
$\Sigma_\star$ there exists a unique solution. Setting the past data to zero eliminates the free
(source-free) sector and yields the unique solution that coincides with the retarded sourced
solution (\ref{eq:Sciama_retarded_rule_SIII}).}.

Therefore, it is useful to make explicit how the inertial calibration responds at linear order
and how this connects to the operational (nonrelativistic) definition of inertial mass.
The appearance of $\sqrt{\mathcal{I}_1}$ in the worldline action follows directly from the fact
that matter is minimally coupled to the physical metric $\tilde g_{\mu\nu}$, while we wish to
express the same dynamics in invariant gravitational units. Using the invariant relation
$\tilde g_{\mu\nu}=\mathcal{I}_1\,\hat g_{\mu\nu}$ (cf.\ Eq.~(\ref{eq:tildeg_in_terms_of_I1_ghat})),
the corresponding proper times are related by
\begin{eqnarray}
d\tilde\tau^2=-\tilde g_{\mu\nu}dx^\mu dx^\nu
=
-\mathcal{I}_1\,\hat g_{\mu\nu}dx^\mu dx^\nu
=\mathcal{I}_1\,d\hat\tau^2
\qquad\Rightarrow\qquad
d\tilde\tau=\sqrt{\mathcal{I}_1}\,d\hat\tau,
\label{eq:dtau_relation_recall}
\end{eqnarray}
for timelike worldlines.

Starting from the universal point-particle action in the physical (matter) frame,
\begin{eqnarray}
S_A=-m_{0A}\int d\tilde\tau,
\end{eqnarray}
where $m_{0A}$ denotes the constant rest-mass parameter of body $A$ and $d\tilde\tau$ is the proper
time measured with the physical metric $\tilde g_{\mu\nu}$, one therefore obtains the
invariant-units form
\begin{eqnarray}
S_A
=
-\,m_{0A}\int d\tilde\tau
=
-\,m_{0A}\int \sqrt{\mathcal{I}_1(\mathcal{I}_3)}\,d\hat\tau,
\label{eq:worldline_action_I1_hat}
\end{eqnarray}
with $d\hat\tau$ the proper time associated with the invariant metric $\hat g_{\mu\nu}$. In this
representation the scalar dependence enters only through the universal factor
$\sqrt{\mathcal{I}_1}$, and the inertial mass scale is identified from the coefficient of the
$\,v^2/2\,$ term in the nonrelativistic expansion of the action.
which makes explicit that the scalar dependence enters only through the invariant factor
$\sqrt{\mathcal{I}_1(\mathcal{I}_3)}$.

To connect with the nonrelativistic notion of inertia, we expand the action in a weak-field,
low-velocity regime using a coordinate time $t$ adapted to the background and writing
\begin{eqnarray}
d\hat\tau
=
dt\,\sqrt{1+2\hat\Psi-v^2}\;,
\qquad
v^2\equiv\delta_{ij}\,\frac{dx^i}{dt}\frac{dx^j}{dt},
\label{eq:dtau_hat_NR_expand}
\end{eqnarray}
where $\hat\Psi$ is the Newtonian potential associated with the metric $\hat g_{\mu\nu}$ in
Newtonian gauge. To first order in $\hat\Psi$ and to second order in $v$ one has
\begin{eqnarray}
\frac{d\hat\tau}{dt}\simeq 1+\hat\Psi-\frac{1}{2}v^2.
\label{eq:dtau_over_dt_hat_NR}
\end{eqnarray}
Substituting (\ref{eq:dtau_over_dt_hat_NR}) into (\ref{eq:worldline_action_I1_hat}) yields the
nonrelativistic action
\begin{eqnarray}
S_A
&\simeq&
\int dt\left[
-\;m_{0A}\sqrt{\mathcal{I}_1(\mathcal{I}_3)}
+\frac{1}{2}\,m_{0A}\sqrt{\mathcal{I}_1(\mathcal{I}_3)}\,v^2
-\;m_{0A}\sqrt{\mathcal{I}_1(\mathcal{I}_3)}\,\hat\Psi
\right].
\label{eq:NR_action_hat_units}
\end{eqnarray}
The constant term is the rest-energy contribution in invariant units, the $v^2/2$ term defines the
inertial mass, and the $\hat\Psi$ term describes coupling to the Newtonian gravitational potential.
Accordingly, the inertial mass measured in invariant gravitational units is
\begin{eqnarray}
m_{I A}(\mathcal{I}_3)\equiv m_{0A}\sqrt{\mathcal{I}_1(\mathcal{I}_3)}.
\label{eq:mI_def_operational}
\end{eqnarray}

When local deviations from a homogeneous cosmological background are small, it is natural to
linearize $\mathcal{I}_1(\mathcal{I}_3)$ about the background value $\bar{\mathcal{I}}_3$. Writing
$\mathcal{I}_3=\bar{\mathcal{I}}_3+\delta\mathcal{I}_3$ with
$|\delta\mathcal{I}_3|\ll |\bar{\mathcal{I}}_3|$, we expand
\begin{eqnarray}
\mathcal{I}_1(\bar{\mathcal{I}}_3+\delta\mathcal{I}_3)
=
\bar{\mathcal{I}}_1
+
\left.\frac{d\mathcal{I}_1}{d\mathcal{I}_3}\right|_{\bar{\mathcal{I}}_3}\,
\delta\mathcal{I}_3
+\cdots,
\label{eq:I1_linearized_SIII}
\end{eqnarray}
with $\bar{\mathcal{I}}_1\equiv \mathcal{I}_1(\bar{\mathcal{I}}_3)$. Combining
(\ref{eq:mI_def_operational}) with (\ref{eq:I1_linearized_SIII}) then gives the linear response of the
inertial calibration,
\begin{eqnarray}
\frac{m_{I A}}{m_{0A}}
=
\sqrt{\mathcal{I}_1}
\simeq
\sqrt{\bar{\mathcal{I}}_1}\left[
1+\frac{1}{2}
\left.\frac{d\ln\mathcal{I}_1}{d\mathcal{I}_3}\right|_{\bar{\mathcal{I}}_3}\,
\delta\mathcal{I}_3
\right],
\label{eq:mI_linearized_response}
\end{eqnarray}
showing explicitly that $\sqrt{\bar{\mathcal{I}}_1}$ sets the dominant background normalization,
while $\delta\mathcal{I}_3$ controls the (causally sourced) spacetime-dependent variations around it.

Only after stating the causal selection rule invariantly do we choose a convenient
representative for explicit cosmological computations. For spatially flat FLRW it is natural
to work in the representative in which matter is minimally coupled to the cosmological metric,
so that the standard FLRW stress-energy conservation law holds with respect to the same metric
used in the line element. We therefore adopt the Sciama matter frame
\begin{eqnarray}
\alpha(\Phi)=0,
\qquad
\tilde g_{\mu\nu}=g_{\mu\nu},
\label{eq:alpha0_choice_SIII}\label{eq:sciama_alpha0}
\end{eqnarray}
which is simply a representative choice (not an additional physical postulate) made for
computational transparency in the FLRW example. In this representative,
\begin{eqnarray}
\mathcal{I}_1(\Phi)=\frac{1}{A(\Phi)},
\label{eq:I1_alpha0_SIII}\label{eq:I1_sciama_alpha0}
\end{eqnarray}
and the inertial scale in invariant gravitational units reads
\begin{eqnarray}
m_{IA}(\Phi)=m_{0A}\sqrt{\mathcal{I}_1(\Phi)}=\frac{m_{0A}}{\sqrt{A(\Phi)}}.
\label{eq:mI_alpha0_SIII}
\end{eqnarray}

Thus, once the causal prescription fixes the cosmological background scalar, it fixes the
relative normalization between inertial scales and the gravitational sector through $A(\bar\Phi)$.

We now specialize to a spatially flat FLRW background in the physical metric $g_{\mu\nu}=\tilde g_{\mu\nu}$,
\begin{eqnarray}
ds^2=-dt^2+a(t)^2\,\delta_{ij}\,dx^i dx^j,
\label{eq:FLRW_metric_SIII}
\end{eqnarray}
and a perfect fluid
\begin{eqnarray}
T_{\mu\nu}=(\rho+p)\,u_\mu u_\nu + p\,g_{\mu\nu},
\qquad
u^\mu=(1,0,0,0),
\qquad
T=-\rho+3p.
\label{eq:perfect_fluid_SIII}
\end{eqnarray}
For dust ($p=0$) one has $T=-\rho_m$.

Since $\hat g_{\mu\nu}=A(\Phi)g_{\mu\nu}$, in four dimensions
\begin{eqnarray}
\sqrt{-\hat g}=A(\Phi)^2\sqrt{-g}=A(\Phi)^2 a^3,
\qquad
\hat g^{tt}=\frac{g^{tt}}{A(\Phi)}=-\frac{1}{A(\Phi)}.
\label{eq:hat_identities_SIII}
\end{eqnarray}
For a homogeneous background $\mathcal{I}_3=\bar{\mathcal{I}}_3(t)$,
\begin{eqnarray}
\hat\square\,\bar{\mathcal{I}}_3
=
\frac{1}{\sqrt{-\hat g}}\partial_t\!\left(\sqrt{-\hat g}\,\hat g^{tt}\,\dot{\bar{\mathcal{I}}}_3\right)
=
-\frac{1}{A}\left[\ddot{\bar{\mathcal{I}}}_3+\left(3H+\frac{\dot A}{A}\right)\dot{\bar{\mathcal{I}}}_3\right],
\label{eq:hatbox_FLRW_SIII}
\end{eqnarray}
where $H=\dot a/a$ and $A=A(\bar\Phi(t))$. Substituting into
(\ref{eq:I3_invariant_eom_SIII}) and using $\hat T=T/A^2$ yields
\begin{eqnarray}
\ddot{\bar{\mathcal{I}}}_3
+
\left(3H+\frac{\dot A}{A}\right)\dot{\bar{\mathcal{I}}}_3
+
\frac{A}{2}\,\frac{d\mathcal{I}_2}{d\mathcal{I}_3}\Big|_{\bar{\mathcal{I}}_3}
=
\frac{A}{4}\,\frac{d\ln\mathcal{I}_1}{d\mathcal{I}_3}\Big|_{\bar{\mathcal{I}}_3}\,\hat T.
\label{eq:I3_FLRW_general_SIII}
\end{eqnarray}
In the Sciama matter frame $\ln\mathcal{I}_1=-\ln A$ and for dust $\hat T=-\rho_m/A^2$, hence
\begin{eqnarray}
\ddot{\bar{\mathcal{I}}}_3
+
\left(3H+\frac{\dot A}{A}\right)\dot{\bar{\mathcal{I}}}_3
+
\frac{A}{2}\,\frac{d\mathcal{I}_2}{d\mathcal{I}_3}\Big|_{\bar{\mathcal{I}}_3}
=
\frac{1}{4A}\,
\frac{d\ln A}{d\mathcal{I}_3}\Big|_{\bar{\mathcal{I}}_3}\,
\rho_m.
\label{eq:I3_FLRW_alpha0_dust_SIII}
\end{eqnarray}

We now explain precisely the approximation behind the ``massless Hubble-damped'' reduction.
Equation (\ref{eq:I3_FLRW_alpha0_dust_SIII}) is a second-order driven equation with Hubble friction.
In many scalar--tensor scenarios relevant for cosmology and local tests, the scalar is light on
Hubble scales and the couplings vary slowly over a Hubble time. Concretely, we assume the regime
\begin{eqnarray}
\left|\frac{A}{2}\frac{d\mathcal{I}_2}{d\mathcal{I}_3}\right|\ll H^2\,|\bar{\mathcal{I}}_3|,
\qquad
\left|\frac{\dot A}{A}\right|\ll H,
\label{eq:light_slow_conditions_SIII}
\end{eqnarray}
so that the effective mass term and the extra friction term $\dot A/A$ do not compete with the
dominant Hubble damping. Under these conditions, (\ref{eq:I3_FLRW_alpha0_dust_SIII}) reduces to
\begin{eqnarray}
\ddot{\bar{\mathcal{I}}}_3 + 3H\dot{\bar{\mathcal{I}}}_3 \simeq S(t),
\qquad
S(t)\equiv
\frac{1}{4A}\left(\frac{d\ln A}{d\mathcal{I}_3}\right)\rho_m,
\label{eq:I3_massless_damped}
\end{eqnarray}
where $d\ln A/d\mathcal{I}_3$ is evaluated on $\bar{\mathcal{I}}_3(t)$.
Equation (\ref{eq:I3_massless_damped}) is equivalent to
\begin{eqnarray}
\frac{d}{dt}\!\left(a^3\,\dot{\bar{\mathcal{I}}}_3\right)=a^3\,S(t),
\label{eq:I3_total_derivative_kernel}
\end{eqnarray}
and imposing the Sciama--Mach condition (no independent homogeneous scalar data) as vanishing
past data at $t_\star$ yields the explicit retarded integral solution
\begin{eqnarray}
\bar{\mathcal{I}}_3(t)
=
\int_{t_\star}^{t} dt'\;K(t,t')\,S(t'),
\qquad
K(t,t')\equiv a^3(t')\int_{t'}^{t}\frac{dt''}{a^3(t'')}.
\label{eq:FLRW_retarded_kernel_solution}
\end{eqnarray}
Thus, in homogeneous FLRW the four-dimensional retarded prescription collapses to a temporal
convolution with a kernel $K(t,t')$ determined by the expansion history.

For power-law expansion $a(t)\propto t^p$ (radiation $p=\tfrac12$, matter $p=\tfrac23$),
\begin{eqnarray}
K(t,t')
=
\frac{t' - t'^{\,3p}\,t^{\,1-3p}}{1-3p}\,\Theta(t-t')
\qquad (p\neq 1/3),
\label{eq:kernel_powerlaw}
\end{eqnarray}
and in particular,
\begin{eqnarray}
K_{\rm rad}(t,t')&=&2\left[t'-(t'^{3/2}/t^{1/2})\right]\Theta(t-t'),
\nonumber\\
K_{\rm mat}(t,t')&=&\left[t'-(t'^2/t)\right]\Theta(t-t'),
\nonumber\\
K_{\rm dS}(t,t')&=&\frac{1}{3H}\left(1-e^{-3H(t-t')}\right)\Theta(t-t').
\label{eq:kernel_examples}
\end{eqnarray}
These explicit kernels show that the scalar response at time $t$ is accumulated from the
source over the only available causal timescale in homogeneous FLRW, namely $\Delta t\sim H^{-1}$
(up to order-unity factors encoded in $K$).

This immediately explains the scaling estimate that plays the role of the FLRW analogue of
Sciama's cosmic-potential argument. A second-order equation with Hubble friction,
$\ddot X+3H\dot X\simeq S$, yields parametrically $\dot X\sim S/H$ and hence $X\sim S/H^2$ over a
Hubble time. Applying this to (\ref{eq:I3_massless_damped}) gives
\begin{eqnarray}
\Delta\bar{\mathcal{I}}_3
\sim
S(t)\,H^{-2}
\sim
\frac{\rho_m}{A}\,H^{-2},
\label{eq:Sciama_scaling_SIII}
\end{eqnarray}
where the slowly varying factor $(d\ln A/d\mathcal{I}_3)$ contributes only an order-unity
coefficient in the regime (\ref{eq:light_slow_conditions_SIII}). In a spatially flat universe
it is natural to relate $\rho_m H^{-2}$ to the matter content inside the Hubble region.
Defining $R_H\sim H^{-1}$ and the enclosed matter mass scale
\begin{eqnarray}
M_H(t)\equiv \frac{4\pi}{3}\,\rho_m(t)\,R_H(t)^3,
\label{eq:Hubble_mass_def}
\end{eqnarray}
one has the identity
\begin{eqnarray}
\rho_m\,R_H^2 \;=\; \frac{3}{4\pi}\,\frac{M_H}{R_H}.
\label{eq:rhoRH2_MoverR}
\end{eqnarray}
Therefore, in homogeneous FLRW the causal sourcing implies that $\bar{\mathcal{I}}_3$ (and hence
$\mathcal{I}_1$ through $\mathcal{I}_1(\mathcal{I}_3)$) responds to a quantity proportional to
$M_H/R_H$, which is the natural FLRW analogue of Sciama's heuristic ``cosmic potential'' scaling.

In this operational sense, the Machian content is carried by the causally selected scalar
configuration: the retarded selection rule fixes $\bar{\mathcal{I}}_3(t)$ from the matter history
in the past light cone of an expanding spatially flat FLRW universe, which fixes $\mathcal{I}_1$
and therefore fixes the relative normalization between local inertial scales and the gravitational
sector.

\section{The weak equivalence principle and the E\"otv\"os parameter}
\label{sec:wep_eotvos}

In the present Sciama-Mach scalar--tensor sector the matter coupling has already been fixed in
Sec.~\ref{sec:sciama_mach} by the universal point-particle action written in invariant gravitational
units, Eq.~(\ref{eq:worldline_action_I1_hat}), and its nonrelativistic reduction,
Eq.~(\ref{eq:NR_action_hat_units}). In particular, the inertial calibration is
\begin{eqnarray}
m_{IA}(\mathcal{I}_3)=m_{0A}\sqrt{\mathcal{I}_1(\mathcal{I}_3)},
\label{eq:mI_recall_IV}
\end{eqnarray}
as given in Eq.~(\ref{eq:mI_def_operational}), and the linear response around a background selected
by the retarded Sciama rule (\ref{eq:Sciama_retarded_rule_SIII})--(\ref{eq:Sciama_free_zero}) is
summarized by Eq.~(\ref{eq:mI_linearized_response}). We now compute explicitly the E\"otv\"os
parameter, emphasizing that for \emph{structureless} test bodies it vanishes identically, while the
only generic opening for $\eta\neq 0$ is the Nordtvedt (self-gravity) channel.

Starting from the nonrelativistic Lagrangian that follows directly from
Eq.~(\ref{eq:NR_action_hat_units}) (dropping the irrelevant constant rest-energy term),
\begin{eqnarray}
L_A
\simeq
\frac{1}{2}\,m_{IA}(\mathcal{I}_3)\,v^2
-\;m_{IA}(\mathcal{I}_3)\,\hat\Psi,
\label{eq:L_NR_start_IV}
\end{eqnarray}
the Euler--Lagrange equation gives
\begin{eqnarray}
\frac{d}{dt}\!\left(\frac{\partial L_A}{\partial v^i}\right)-\frac{\partial L_A}{\partial x^i}=0
\qquad\Rightarrow\qquad
\frac{d}{dt}\!\left(m_{IA} v^i\right)
=
-\;m_{IA}\,\partial_i\hat\Psi
-\;(\partial_i m_{IA})\,\hat\Psi
+\;\frac{1}{2}\,(\partial_i m_{IA})\,v^2.
\label{eq:EOM_exact_NR_IV}
\end{eqnarray}
In a free-fall/slow-motion regime ($v^2\ll 1$) and weak field ($|\hat\Psi|\ll 1$), the last two terms
are subleading compared to $m_{IA}\partial_i\hat\Psi$, and one obtains
\begin{eqnarray}
\frac{d}{dt}\!\left(m_{IA} v^i\right)\simeq -\,m_{IA}\,\partial_i\hat\Psi.
\label{eq:EOM_leading_NR_IV}
\end{eqnarray}
Expanding the time derivative yields
\begin{eqnarray}
m_{IA} a_A^i + \dot m_{IA}\,v^i \simeq -\,m_{IA}\,\partial_i\hat\Psi,
\label{eq:EOM_expand_IV}
\end{eqnarray}
and neglecting the velocity-suppressed term $\dot m_{IA} v^i$ gives the standard Newtonian form
\begin{eqnarray}
a_A^i \simeq -\,\partial_i\hat\Psi.
\label{eq:Newtonian_acc_IV}
\end{eqnarray}
At the same level of approximation, one may keep the leading ``fifth-force'' correction arising
from the spatial dependence of $m_{IA}$ itself by rewriting (\ref{eq:EOM_leading_NR_IV}) as
\begin{eqnarray}
a_A^i
=
-\,\partial_i\hat\Psi
-\;\partial_i\ln m_{IA}
+\;\mathcal{O}(\hat\Psi\,\partial \ln m_{IA},\,v^2\partial \ln m_{IA}).
\label{eq:acc_with_massgrad_IV}
\end{eqnarray}
Using Eq.~(\ref{eq:mI_recall_IV}),
\begin{eqnarray}
\partial_i\ln m_{IA}
=
\frac{1}{2}\,\partial_i\ln\mathcal{I}_1
=
\frac{1}{2}\,
\left(\frac{d\ln\mathcal{I}_1}{d\mathcal{I}_3}\right)\partial_i\mathcal{I}_3.
\label{eq:grad_ln_mI_universal_IV}
\end{eqnarray}
This is the crucial point: the coefficient multiplying $\partial_i\mathcal{I}_3$ is \emph{universal}
and therefore independent of composition labels. The Sciama--Mach postulate determines which
$\delta\mathcal{I}_3$ is physically realized (retarded sourcing), but it does not introduce
body-dependent couplings for structureless test particles.

Consider now two structureless test bodies $A$ and $B$ released in the same external fields
$(\hat\Psi,\mathcal{I}_3)$. From Eqs.~(\ref{eq:acc_with_massgrad_IV})--(\ref{eq:grad_ln_mI_universal_IV}),
\begin{eqnarray}
a_A^i-a_B^i=0,
\label{eq:deltaa_zero_structureless_IV}
\end{eqnarray}
and the E\"otv\"os parameter \cite{Will2014,Touboul2022_MICROSCOPE,Ross2025WEP,Wagner2012WEP},
\begin{eqnarray}
\eta_{AB}\equiv 2\,\frac{|a_A-a_B|}{|a_A+a_B|}
\label{eq:Eotvos_def_IV}
\end{eqnarray}
vanishes identically,
\begin{eqnarray}
\eta_{AB}=0,
\qquad
\text{(structureless test bodies, universal matter coupling).}
\label{eq:Eotvos_zero_IV}
\end{eqnarray}
Hence, within the universally coupled Sciama--Mach sector, the weak equivalence principle holds for
all bodies whose internal structure does not contribute appreciably to their total mass.

The only generic possibility for a nonzero $\eta_{AB}$ is associated with \emph{self-gravitating}
bodies (Nordtvedt effect), for which the total mass depends on the ambient scalar configuration via
the gravitational binding energy. A convenient invariant parametrization introduces the
sensitivity
\begin{eqnarray}
s_A
\;\equiv\;
-\,\left.\frac{\partial\ln m_A}{\partial\ln \mathcal{I}_1}\right|_{\bar{\mathcal{I}}_3},
\label{eq:sensitivity_def_IV}
\end{eqnarray}
so that the scalar-induced acceleration becomes body-dependent at leading order. In that case one
may write schematically
\begin{eqnarray}
\partial_i\ln m_A
\simeq
\frac{1}{2}\,(1-2s_A)\,\partial_i\ln\mathcal{I}_1,
\label{eq:grad_ln_m_selfgrav_IV}
\end{eqnarray}
which replaces the universal structureless result (\ref{eq:grad_ln_mI_universal_IV}). Substituting
into (\ref{eq:acc_with_massgrad_IV}) yields
\begin{eqnarray}
a_A^i-a_B^i
\simeq
-\,(s_A-s_B)\,\partial_i\ln\mathcal{I}_1
=
-\,(s_A-s_B)\left(\frac{d\ln\mathcal{I}_1}{d\mathcal{I}_3}\right)\partial_i\delta\mathcal{I}_3,
\label{eq:deltaa_Nordtvedt_IV}
\end{eqnarray}
so that the E\"otv\"os parameter $\eta_{AB}$ can depart from zero precisely when $s_A\neq s_B$.
Therefore, in the present framework-universal conformal coupling of matter together with the
Sciama-Mach retarded selection rule-the universality of free fall is preserved for structureless
test bodies at leading order, and any generic deviations from the equivalence principle, if
present at all, are restricted to the (Nordtvedt) self-gravity.

\section{Conclusions}
\label{sec:conclusions}

We have presented a scalar realization of Sciama’s Machian programme in a cosmological setting within the Bergmann--Wagoner class of scalar--tensor theories. Starting from a general action with universal (conformal) coupling of matter to a single physical metric, we derived the corresponding field equations and recast the theory in terms of a frame-invariant set of variables. This invariant formulation provides a representation-independent language in which Machian requirements can be stated as genuine physical restrictions on the admissible solution space, rather than as frame-dependent statements about the scalar field or the nonminimal coupling function.

Within this framework, Sciama’s causal postulate was implemented as a \emph{selection principle} rather than as a modification of the local dynamics. Concretely, among the solutions of the same hyperbolic scalar equation, we retain only those configurations that are generated by the matter distribution in the causal past, i.e.\ the retarded response, with any independent source-free (homogeneous) contribution removed. In a spatially flat FLRW background and in the regime where the scalar is light and slowly varying on Hubble scales, this prescription collapses to an explicit temporal kernel: the cosmological scalar evolution at a given time is determined by a causal time integral of the matter source. This result yields a transparent FLRW analogue of Sciama’s scaling argument: the dominant response is controlled by the matter content within the Hubble region, so that the cosmological history fixes the inertial normalization in a concrete and covariant manner.

A central outcome is that the same causally selected background determines the inertial calibration of matter through the universal coupling to the physical metric. In the test-body limit, the inertial mass scale is set by a single universal factor (equivalently, by $\sqrt{\mathcal I_1}$ in invariant variables), so that all structureless bodies respond identically to a given external configuration. Consequently, the weak equivalence principle is preserved at leading order: the differential acceleration between two structureless test bodies vanishes and the E\"otv\"os parameter is zero. Any generic departure from universality can arise only when gravitational binding energy contributes non-negligibly to the total mass, i.e.\ through the self-gravity (Nordtvedt), where body-dependent sensitivities can induce composition-dependent scalar responses.

Thus, natural extensions include going beyond the light-field approximation to quantify how an effective scalar mass and slowly varying couplings modify the causal kernel and its scaling across cosmological eras, constructing explicit realizations of induced-gravity branches and screening-compatible models and mapping their global solution space (including criteria excluding de Sitter-like inertial vacua), and deriving observable predictions through invariant post-Newtonian, strong-field, and cosmological-perturbation analyses. It is also worth investigating whether broader selection principles, including anthropic reasoning and observational selection effects, can further constrain admissible cosmological histories and the resulting inertial calibration.

\section*{Acknowledgments} 

I would like to thank Maria Margarita and Miguel Amado for their continuous inspiration and support, and the Foundation for Research Support of Espírito Santo (FAPES) for the partial support for the present work.


\begin{thebibliography}{99}

\bibitem{LeibnizClarke1715}
G.~W.~Leibniz and S.~Clarke,
\textit{The Leibniz--Clarke Correspondence},
ed. H.~G.~Alexander
(Manchester Univ. Press, Manchester, 1956).

\bibitem{Berkeley1721}
G.~Berkeley,
\textit{De Motu} (1721),
in \textit{The Works of George Berkeley}, Vol.~IV,
ed. A.~A.~Luce
(Nelson, London, 1948).

\bibitem{Berkeley1710}
G.~Berkeley,
\textit{A Treatise Concerning the Principles of Human Knowledge}
(Dublin, 1710).

\bibitem{Mach1883}
E.~Mach,
\textit{Die Mechanik in ihrer Entwicklung historisch-kritisch dargestellt}
(F.~A.~Brockhaus, Leipzig, 1883);
English transl. \textit{The Science of Mechanics}
(Open Court, La Salle, IL, 1919).

\bibitem{Einstein1916GR}
A.~Einstein,
Ann.\ Phys.\ (Berlin) \textbf{49}, 769 (1916),
doi:10.1002/andp.19163540702.

\bibitem{Einstein1916Mach}
A.~Einstein,
Phys.\ Z.\ \textbf{17}, 101 (1916).

\bibitem{Einstein1918}
A.~Einstein,
Ann.\ Phys.\ (Berlin) \textbf{55}, 241 (1918),
doi:10.1002/andp.19183600402.

\bibitem{SEPGenRelEarly}
C.~W.~Misner, K.~S.~Thorne, and J.~A.~Wheeler,
\textit{Gravitation}
(W.~H.~Freeman, San Francisco, 1973).

\bibitem{BondiSamuel1964}
H.~Bondi and J.~Samuel,
``The Lense--Thirring Effect and Mach's Principle,''
Phys.\ Lett.\ \textbf{13}, 132--134 (1964).


\bibitem{Raine1981}
D.~J.~Raine,
Rep.\ Prog.\ Phys.\ \textbf{44}, 1151 (1981),
doi:10.1088/0034-4885/44/11/001.

\bibitem{BarbourBertotti1982}
J.~B.~Barbour and B.~Bertotti,
Proc.\ R.\ Soc.\ Lond.\ A \textbf{382}, 295 (1982),
doi:10.1098/rspa.1982.0102.

\bibitem{BarbourPfister1995}
J.~Barbour and H.~Pfister (eds.),
\textit{Mach's Principle: From Newton's Bucket to Quantum Gravity},
Einstein Studies, Vol.~6
(Birkh\"auser, Boston, 1995).

\bibitem{Ellis2001}
G.~F.~R.~Ellis,
New Astron.\ Rev.\ \textbf{46}, 645 (2002),
doi:10.1016/S1387-6473(02)00234-8,
arXiv:gr-qc/0102017.

\bibitem{LichteneggerMashhoon2004}
H.~I.~M.~Lichtenegger and B.~Mashhoon,
in \textit{The Measurement of Gravitomagnetism},
ed. L.~Iorio \textit{et al.}
(Nova Science, New York, 2004),
arXiv:physics/0407078.

\bibitem{Gine2006}
J.~Gin\'e,
Int.\ J.\ Theor.\ Phys.\ \textbf{45}, 457 (2006).

\bibitem{Barbour2010Definition}
J.~Barbour,
Found.\ Phys.\ \textbf{40}, 1263 (2010),
doi:10.1007/s10701-010-9490-7.
Licata2014, Kragh2012
]
\bibitem{Licata2014}
I.~Licata and L.~Chiatti,
Electron.\ J.\ Theor.\ Phys.\ \textbf{11}, 41 (2014).

\bibitem{Kragh2012}
H.~Kragh,Fay2024Sciama
``Mach's Principle and the Origin of General Relativity,''
in \textit{The Genesis of General Relativity},
Vol.~3, eds.\ J.~Renn and M.~Schemmel,
Springer, Dordrecht (2012), pp.~319--345.


\bibitem{Fay2024Sciama}
N.~Sultana and D.~Kazanas,
Int.\ J.\ Mod.\ Phys.\ D \textbf{20}, 1205 (2011),
doi:10.1142/S0218271811019384,
arXiv:1104.1306.

\bibitem{Fay2024MachHypotheses}
J.~Fay,
Stud.\ Hist.\ Philos.\ Sci.\ \textbf{103}, 58 (2024),
doi:10.1016/j.shpsa.2023.09.006.




\bibitem{BransDicke1961}
C.~Brans and R.~H.~Dicke,
Phys.\ Rev.\ \textbf{124}, 925 (1961),
doi:10.1103/PhysRev.124.925.


\bibitem{Sciama1953}
D.~W.~Sciama,
Mon.\ Not.\ R.\ Astron.\ Soc.\ \textbf{113}, 34 (1953),
doi:10.1093/mnras/113.1.34.

\bibitem{Sciama1959}
D.~W.~Sciama,
\textit{The Unity of the Universe}
(Faber and Faber, London, 1959).

\bibitem{Sciama1964}
D.~W.~Sciama,
Rev.\ Mod.\ Phys.\ \textbf{36}, 463 (1964),
doi:10.1103/RevModPhys.36.463.







\bibitem{Bergmann1968}
P.~G.~Bergmann,
Int.\ J.\ Theor.\ Phys.\ \textbf{1}, 25 (1968),
doi:10.1007/BF00668828.

\bibitem{Wagoner1970}
R.~V.~Wagoner,
Phys.\ Rev.\ D \textbf{1}, 3209 (1970),
doi:10.1103/PhysRevD.1.3209.



\bibitem{FaraoniBook2004}
V.~Faraoni,
\textit{Cosmology in Scalar-Tensor Gravity}
(Kluwer Academic, Dordrecht, 2004).

\bibitem{FujiiMaeda2003}
Y.~Fujii and K.~Maeda,
\textit{The Scalar-Tensor Theory of Gravitation}
(Cambridge Univ. Press, Cambridge, 2003).



\bibitem{DamourEspositoFarese1996BinaryPulsar}
T.~Damour and G.~Esposito-Far\`ese,
Phys.\ Rev.\ D \textbf{54}, 1474 (1996),
doi:10.1103/PhysRevD.54.1474,
arXiv:gr-qc/9602056.




\bibitem{Wald1984}
R.~M.~Wald,
\textit{General Relativity}
(University of Chicago Press, Chicago, 1984).






\bibitem{JarvKuuskSaalVilson2015Invariant}
L.~J\"arv, P.~Kuusk, M.~Saal, and O.~Vilson,
Phys.\ Rev.\ D \textbf{91}, 024041 (2015),
doi:10.1103/PhysRevD.91.024041,
arXiv:1411.1947.

\bibitem{JarvKuuskSaalVilson2015Transform}
L.~J\"arv, P.~Kuusk, M.~Saal, and O.~Vilson,
Class.\ Quant.\ Grav.\ \textbf{32}, 235013 (2015),
doi:10.1088/0264-9381/32/23/235013,
arXiv:1504.02686.



\bibitem{KaramTamvakis2018FrameInvariant}
A.~Karam, A.~Lykkas, and K.~Tamvakis,
``Frame-invariant approach to higher-dimensional scalar-tensor gravity,''
Phys.\ Rev.\ D \textbf{97}, 124036 (2018),
doi:10.1103/PhysRevD.97.124036.

\bibitem{Dicke1962}
R.~H.~Dicke,
``Mach's principle and invariance under transformation of units,''
Phys.\ Rev.\ \textbf{125}, 2163--2167 (1962),
doi:10.1103/PhysRev.125.2163.

\bibitem{KaramTamvakis2019Inflation}
A.~Karam, A.~Lykkas, and K.~Tamvakis,
``Frame-invariant approach to inflation in scalar--tensor theories,''
Phys.\ Rev.\ D \textbf{99}, 064029 (2019),
doi:10.1103/PhysRevD.99.064029.





\bibitem{Will2014}
C.~M.~Will,
Living Rev.\ Relativ.\ \textbf{17}, 4 (2014),
doi:10.12942/lrr-2014-4.

\bibitem{Touboul2022_MICROSCOPE}
P.~Touboul \textit{et al.},
Phys.\ Rev.\ Lett.\ \textbf{129}, 121102 (2022),
doi:10.1103/PhysRevLett.129.121102.


\bibitem{Ross2025WEP}
M.~P.~Ross, S.~M.~Fleischer, I.~A.~Paulson, P.~Lamb, B.~M.~Iritani, E.~G.~Adelberger, C.~A.~Hagedorn, K.~Venkateswara, C.~Gettings, E.~A.~Shaw, S.~K.~Apple, and J.~H.~Gundlach,
``Test of the Equivalence Principle for Superconductors,''
Phys.\ Rev.\ D \textbf{111}, L021101 (2025),
doi:10.1103/PhysRevD.111.L021101.

\bibitem{Wagner2012WEP}
T.~A.~Wagner, S.~Schlamminger, J.~H.~Gundlach, and E.~G.~Adelberger,
``Torsion-balance tests of the weak equivalence principle,''
Class.\ Quant.\ Grav.\ \textbf{29}, 184002 (2012),
doi:10.1088/0264-9381/29/18/184002.



\end{thebibliography}
\end{document}